\documentclass[final,3p,times]{elsarticle}
\usepackage{lineno,hyperref,xcolor}
\usepackage{caption}
\usepackage{booktabs}
\usepackage{cleveref}
\modulolinenumbers[5]
\biboptions{numbers,sort&compress}
\usepackage{subfigure}

\hypersetup{colorlinks,
	linkcolor=Cerulean,%
	citecolor=Cerulean}

\journal{Journal of \LaTeX\ Templates}

\begin{document}
	\begin{frontmatter}
		
		\title{Population rate coding in recurrent neuronal networks with undetermined-type neurons}
		
		\author[mymainaddress]{Hao Si}
		
		\author[mymainaddress]{Xiaojuan Sun\corref{mycorrespondingauthor}}
		\cortext[mycorrespondingauthor]{Correspondence and requests for materials should be addressed to X. S}
		\ead{sunxiaojuan@bupt.edu.cn}		
		
		\address[mymainaddress]{School of Science, Beijing University of Posts and Telecommunications, Beijing 100876, P.R. China.}	
		
\begin{abstract}
	Neural coding is a key problem in neuroscience, which can promote people's understanding of the mechanism that brain processes information. Among the classical theories of neural coding, the population rate coding has been studied widely in many works. Most computational studies considered the neurons and the corresponding presynaptic synapses as pre-determined excitatory or inhibitory types. According to physiological evidence, however, that the real effect of a synapse is inhibitory or excitatory is determined by the type of the activated receptors. The co-release of excitatory and inhibitory receptors in the same synapse exists widely in the brain. In this paper, we study the population rate coding in recurrent neuronal networks with undetermined neurons and synapses, different from the traditional works, in which one neuron can perform either excitatory or inhibitory effect to the corresponding postsynaptic neurons. We find such neuronal networks can encode the stimuli information in population firing rate well. We find that intermediate recurrent probability together with moderate Inhibitory-Excitatory strength ratio can enhance the encoding performance. Suitable combinations of the previous two parameters with the noise intensity, the excitatory synaptic strength and the synaptic time constant have promoting effects on the performance of population rate coding. Finally, we compare the performance of population rate coding between the traditional (determined) model and ours, and we find that it is rational to consider the co-release of inhibitory and excitatory receptors.
	
\end{abstract}

\begin{keyword}
	\texttt{Neural coding, population rate coding, recurrent neuronal network}
\end{keyword}

\end{frontmatter}


\section{Introduction}

One of the most fundamental question in neuroscience is the problem of neural coding, which is a key to understand the brain. A neural coding is a system of rules and mechanisms by which a signal carries information\cite{deCharms2000}. To our knowledge, generally two types of neural coding are mainly considered including rate coding\cite{Newsome1989,Georgopoulos1993,Shadlen1994,Marsalek1997,Rossum2002,Mazurek2002}and temporal coding\cite{Aertsen1996,Riehle1997,Diesmann1999,Litvak2003,Vogels2005}. 
In the rate code paradigm, neuronal information is carried by the mean firing rate of neurons. In the temporal code paradigm, which is also called synchrony code, neuronal information is represented by the precise spiking timings of neurons. Though the two paradigms are alternative to each other, however, some works thought that the brain may work in a combination of both paradigms\cite{Masuda2002,Masuda2003DualityOR,Masuda2004,Bruno12006,Kumar2010}. Up to now, how neuronal information is coded in the brain cortex is still under debate.

In these neural coding hypothesis, population rate coding is an ingenious hypothesis worth paying attention to. According to the classical rate coding, neurons encode the information in rate coding through representing it in the number of spikes per observation time window(firing rate)\cite{Adrian1928,Kostal2007}. The neurons, however, need a short integration time to estimate the signal\cite{Shadlen1998}. If it is required for every synaptic stage to average in time calculating the firing rate, the rate code is highly impossible for the information coding in the brain in that the brain is a highly efficient machine so its mechanism of information coding is ought to be efficient. The population rate coding was introduced, which calculates the population firing rate instead of the mean firing rate, improving the efficiency of rate coding. The population rate coding is based on the experimental observation with the intensity of the external stimuli\cite{Gerstner2000,Kandel1991}. The population rate coding is capable of faster and more accurate processing since averaging is performed across many fast responding neurons\cite{Brunel1999,Knight2000,Gerstner2000,Dayan2005}. Recently, the population rate coding has been widely investigated in many works\cite{Masuda2003DualityOR,Masuda2005,Rossum2002,Wang2009,Guo2012}.

In most theoretical studies mentioned above, they only considered excitatory neurons or synapses but ignored the recurring properties in neuronal networks. In the real biological neuronal network, there exist not only excitatory neurons or synapses. The dynamics of excitatory-inhibitory (E-I) neuronal networks has been investigated in many works\cite{van1996,Emilio2001,Brunel2000,Carsten2003,Teramae2007}. Activity patterns and spiking dynamics in E-I recurrent networks were systematically studied in these works. What's more, signal coding in recurrent E-I network also attracted much research\cite{Brunel2000,Vogels2005,Kremkow2010,Kumar2008,Kumar2010,Vogels2009,Kremkow2010,Julien2004,Wang2009,Guo2012,Han2015,Xiaojing2017}. In most works, neurons are considered into excitatory and inhibitory neurons and whether the synaptic connections are excitatory or inhibitory is mainly determined by the type of presynaptic neurons. However, according to physiological evidences, the property of the synapses should be determined by the type of activated receptors\cite{Neuron-wiki}. If excitatory receptors are activated, then the synapses are excitatory; Otherwise, they are inhibitory. Shrivastava et al. reviewed that inhibitory receptor GABA co-releases with the excitatory receptor, e.g., glutamate, at the same presynaptic synapse, which suggested that the cross-talk between different types of receptors might be a general phenomenon in the nervous system\cite{Shrivastava2011}. Furthermore, recently Kantamneni reviewed that how GABA receptor activity influences glutamate receptor function and \emph{vice versa}, which concluded that the cross-talk between excitatory and inhibitory receptors plays a key role in the balance between excitatory and inhibitory neurotransmission in brain\cite{Kantamneni2015}. Therefore, inspired by these experimental studies, we construct a novel neuronal network, instead of taking pre-determined excitatory and inhibitory neurons, in which the excitatory and inhibitory synapses grow from the same one neuron with the ratio of 4:1. Meanwhile, for simplicity, that one postsynaptic neuron receives the excitatory and inhibitory synapse from the same presynaptic neuron at the same time is avoided. Unlike the former works, in this paper, we will discuss population rate coding in such neuronal network, in which the excitation or inhibition of synaptic connections is just determined by the type of activated receptors, namely, the same presynaptic neurons might perform excitatory or inhibitory effects on the corresponding postsynaptic neurons. In the following content, we will mainly study the representation of population firing rate in such a recurrent neuronal network.

In this paper, we construct a neuronal network in a more biological manner to study the population rate coding. The contents are arranged as follows. In 'Model and method' section, we introduce the computational model of neurons, synapses, and networks as well as the measurements of the neuronal information coding. In 'Results' section,  the effects of some critical parameters on the population rate coding will be illustrated and discussed. Lastly, we try to compare the performance of population rate coding in the different model and ours. In 'Conclusion' section, we summarize the conclusions of this work and have some discussions about the further works in the future.

\section{Model and Method}	
\subsection{Network Topology}

\begin{figure}[!htbp]
	\centering
	\includegraphics[width=11.5cm]{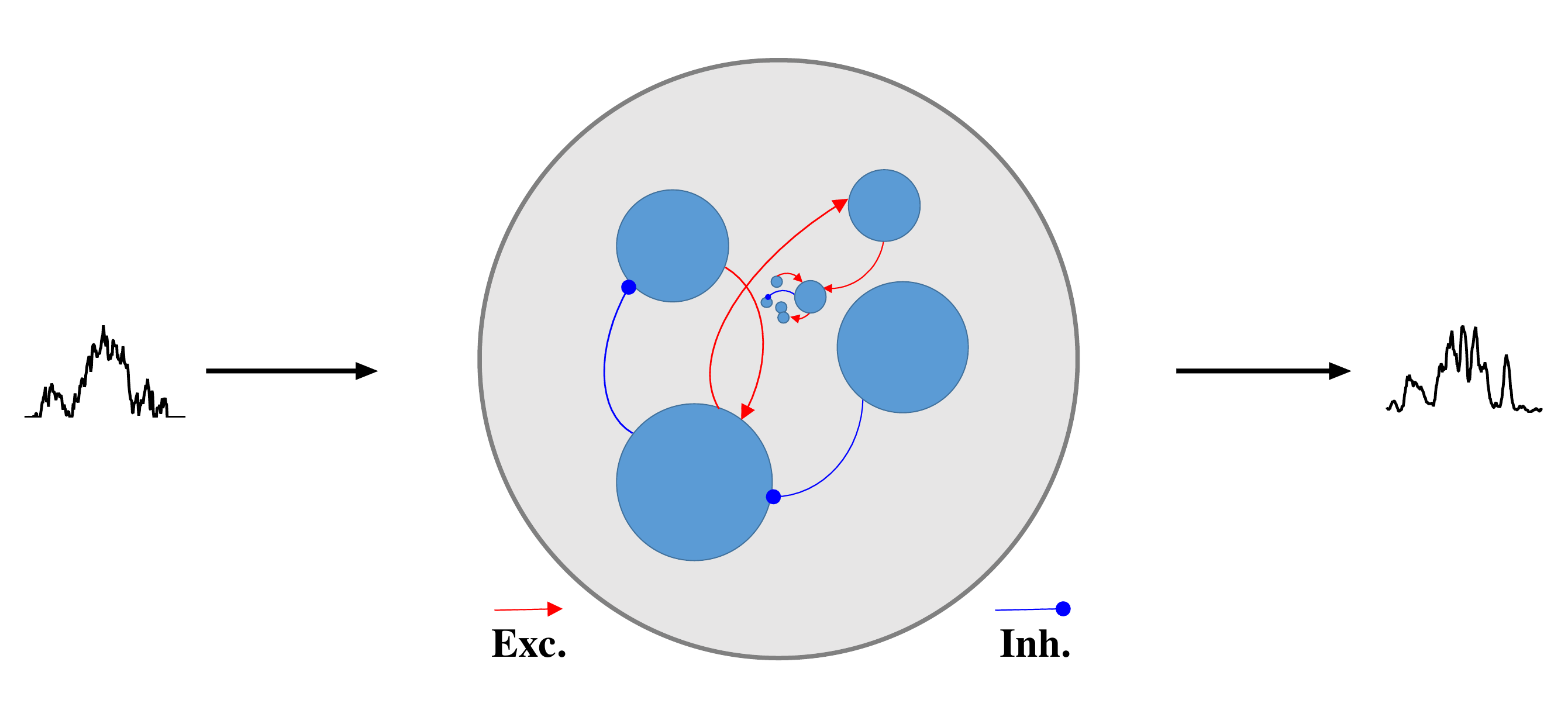}
	\caption{The illustration of the recurrent network. The light blue circles denote the neurons whose total number is $N$. The red arrow lines denote excitatory synaptic connections while the blue lines with a point denote inhibitory synaptic connections. The ratio between the quantity of excitatory and inhibitory synaptic connection in such a network is about 4:1.}
	\label{fig.1}
\end{figure}

The recurrent network constructed in this paper is shown in \ref{fig.1}.
The network consists of $N$ neurons coupled through excitatory and inhibitory synaptic connections. Neurons connect to each other with probability $P_{rc}$ through  recurrent connections, $80\%$ of which are excitatory synaptic connections while $20\%$ are inhibitory, namely, excitatory probability $P_{rc}^E = 0.8$ and inhibitory probability $P_{rc}^I = 0.2$. Notably, excitatory connections could grow from a neuron who also grow inhibitory connections.  Generally, $N=100$  and $P_{rc}=0.12$ otherwise specified.

\subsection{Neurons and Synapses}
The neuron model used in this paper is the Hodgkin-Huxley model\cite{Hodgkin1952} whose dynamics is shown as follows.

\begin{equation}
\left\{
\begin{array}{lr}
C\frac{dV_{i}}{dt}=I^{ext}-g_{Na}m_{i}^{3}h_{i}(V_{i}-V_{Na})-g_{K}n_{i}^{4}(V_{i}-V_{K})-g_{l}(V_{i}-V_{l})+I_{i}^{syn}+D\xi, &  \\
\frac{dm_{i}}{dt}=a_m(V_m)(1-m)-b_m(V_m)m, & \\
\frac{dh_{i}}{dt}=a_h(V_h)(1-h)-b_h(V_h)h, & \\
\frac{dn_{i}}{dt}=a_n(V_n)(1-n)-b_n(V_n)n, & \\
\end{array}
\right.
\label{HH_model}
\end{equation}
where $V_{i}$ denotes the membrane potential of each neuron $i$ $(i=1,\ldots,N)$. $C = 1\, \mu F/cm^2$ is the membrane capacity. $I^{ext}$ denotes the external stimuli injected into all neurons in the network. $m,h,n$ are activation and inactivation gate variables of sodium channels, activation gate variable of potassium channels, respectively. $g_{Na} = 120\,mS/cm^2; g_K = 36\, mS/cm^2$ are the maximal sodium and potassium conductance, respectively. $V_{Na}=50\, mV$, $V_K=-77\, mV$, $V_l=-54.4\, mV$ are the reversal potential of sodium, potassium, leaky currents, respectively. $I_{i}^{syn}$ is the synaptic current of each neuron from the coupled presynaptic neurons. $D\xi$ is the background noise with $D$ denoting the noise intensity and $\xi$ being the white Gaussian noise, whose mean is zero and the standard deviation is 1. The function $a_x(V)$ and $b_x(V)$ with $x = m,n,h$ are given by: $a_m=0.1(V+40)/(1-exp[(-V-40)/10])$, $b_m=4exp[(-V-65)/18]$, $a_h=0.07exp[(-V-65)/20]$, $b_h=1/(1+exp[(-V-35)/10])$, $a_n=0.01(V+55)/(1-exp[(-V-55)/10])$, $b_n=0.125exp[(-V-65)/80]$ from Hansel et al\cite{Hansel1993}.

A neuron's synaptic currents are shown as follows. Synaptic connections here are modeled as a conductance-based model.
\begin{equation}
I_{i}^{syn}=\sum_{1}^{N}G(t)(V(t)-V_{syn}),
\end{equation}
where $N$ denotes the amount of presynaptic neurons coupled with the neuron $(i)$. $V_{syn}$ denotes the reversal potential of synapse with $V_{syn} = 0$ for excitatory connections and $V_{syn} = -75\, mV$ for inhibitory connections. $G(t)$ denotes the synaptic conductance. In detail, the conductance is written as
\begin{equation}
G_x(t)=\left\{
\begin{array}{rcl}
g_x\alpha(t-t_{spk}), & & {t-t_{spk}>0}\\
0, & & {else}\\
\end{array}, \right.
\end{equation}
with  $\alpha(t)=(t/\tau)e^{(-t/\tau)}$. $t_{spk}$ is the spike timings of the presynaptic neurons. $\tau$ is synaptic time constant. For excitatory connections, $\tau^{exc}=0.3\, ms$. For inhibitory connections, we set $\tau^{inh}=0.6\, ms$. If not specified below, the excitatory synaptic strength is set as $g^{exc}=0.102$. The inhibitory synaptic strength $g^{inh}$ is calculated by excitatory one. The ratio between them is
\begin{equation}
R = \frac{g^{inh}}{g^{exc}},
\end{equation}
through which we could control the value of $g^{inh}$ to modulate the degree of inhibition in the population as well as the population firing pattern.
\subsection{Time-varying external input current}
In order to study the rate code, we use a kind of time-varying external input current $I(t)$ which is used ubiquitously in many papers\cite{Rossum2002,Wang2009,Guo2012,Han2015}, also named a half-wave rectified Gaussian noise.
\begin{equation}
I(t)=\left\{
\begin{array}{rcl}
K \eta(t), & & {if\; \eta\geq 0}\\
0, & & {if\; \eta <0}\\
\end{array}, \right.
\end{equation}
with $K=15$ denoting the modulation strength. $\eta(t)$ is an Ornstein-Uhlenbeck process whose dynamics equation is
\begin{equation}
\tau _c\frac{d\eta(t)}{dt}=-\eta(t)+\sqrt{2A}\xi(t),
\end{equation}
where $\xi(t)$ is a Gaussian white noise, correlation time $\tau _c$ is set as 80 ms. If not specified, the external input current intensity is set as $A = 200$.

\subsection{Methods}
\subsubsection{Population firing rate}
The population firing rate is used to represent the signal, which is calculated as
\begin{equation}
p(t)=\frac{N(\Delta t)}{\Delta t},
\end{equation}
where $\Delta t$ is a given short time interval centered at time $t$, whose value here is taken as 1 $ms$, $N(\Delta t)$ is the total amount of spikes in the neuronal population during the given time interval.
\subsubsection{Encoding quality}
The correlation coefficient $C(\tau)$ of input $s(t)$ and population rate $p(t)$ in target layer $i$ is introduced to quantify how well the input is encoded and propagated by the network. $C(\tau)$ is calculated as\cite{Vogels2005}
\begin{equation}
C(\tau )=\frac{\left \langle [s(t)-\bar{s}][p(t+\tau )-\bar{p}]\right \rangle_{t}}{\sqrt{\left \langle [s(t)-\bar{s}]^{2} \right \rangle_{t}\left \langle [p(t+\tau )-\bar{p}]^{2} \right \rangle_{t}}},
\end{equation}
where $p(t)$ is the population firing rate in a 10 $ms$ time window sliding with a step of 1 $ms$.		
Here we term the maximum of the correlation coefficients as encoding fidelity, ${Q=max\{(C(\tau)\}}$, through which we could know how much information is captured by population firing rate.
\subsubsection{Synchrony degree }
The synchrony degree is used to quantify how synchronously the neurons in the network generate spikes, by which we could determine the local state of neuronal population and do more analysis. The synchrony degree $\mathit{S}$ is quantified by the population coherence measure $\mathit{S}$. $\mathit{S}$ is calculated as\cite{Wang1996} 	
\begin{equation}
S=\frac{1}{N(N-1)}\sum_{i=1}^{N}\sum_{j=1,j\neq i}^{ }k_{i,j}(\Delta t),
\end{equation}
where $\mathit{N}$ is the size of a subnetwork. $\mathit{k}$$_{i,j}$ is the coherence measure of a neuron pair calculated as
\begin{equation}
k_{i,j}=\frac{\sum_{l=1}^{T/\Delta t}X(l)Y(l)}{\sqrt{\sum_{l=1}^{T/\Delta t}X(l)\sum_{l=1}^{T/\Delta t}Y(l)}},
\end{equation}
where $\mathit{X,Y}$ are the discretizing action potential trains of neuron pair respectively. In detail, one would divide the two time trains whose interval is $\mathit{T}$ into small bins of $\tau$, then write the trains as
\begin{center}
	$	X(l),Y(l)=\left\{
	\begin{array}{rcl}
	1, & & {if \; spike}\\
	0, & & {else}\\
	\end{array}, \right.$
\end{center}
where ${l=1,\ldots,T/\Delta t}$.

\section{Results}
In this section, the simulation results and relevant analysis are presented. We mainly pay attention to discussing how stochastic stimuli can be represented with the population firing rate in the studied neuronal network and how to improve the encoding quality. The key parameters we investigate include I/E strength ratio $R$, recurrent probability $P_{rc}$, noise intensity $D$, excitatory strength $g^{exc}$, and the synaptic time constant $\tau^{exc},\tau^{inh}$.
Finally, we compare the encoding quality in our considered network with the neuronal network with determined neuron types in the previous studies.

\subsection{I/E strength ratio and recurrent probability}
\begin{figure}[!htbp]
	\centering
	\subfigure[]{
		\includegraphics[width=5.4cm]{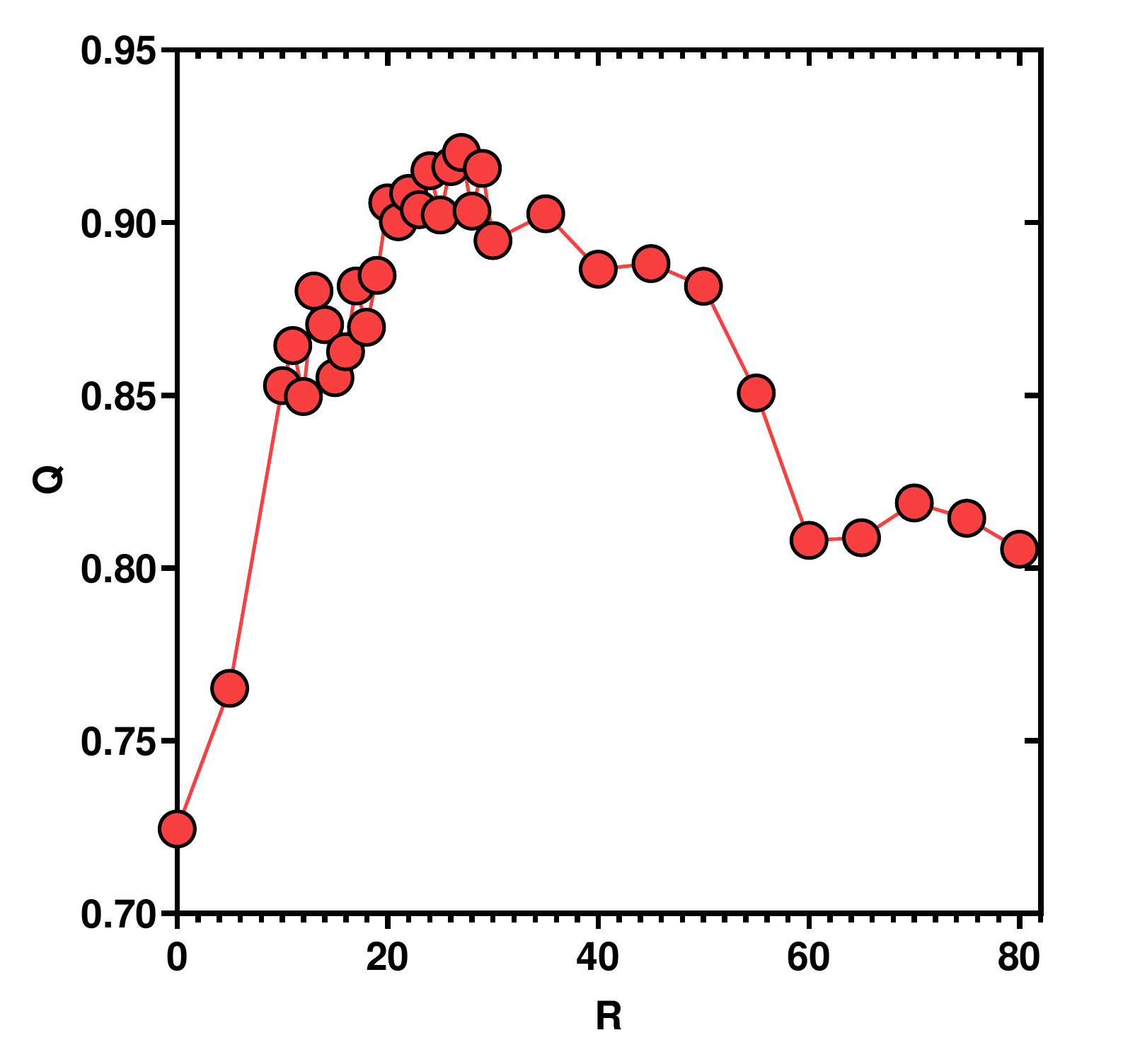}}
	\subfigure[]{
		\includegraphics[width=6.0cm]{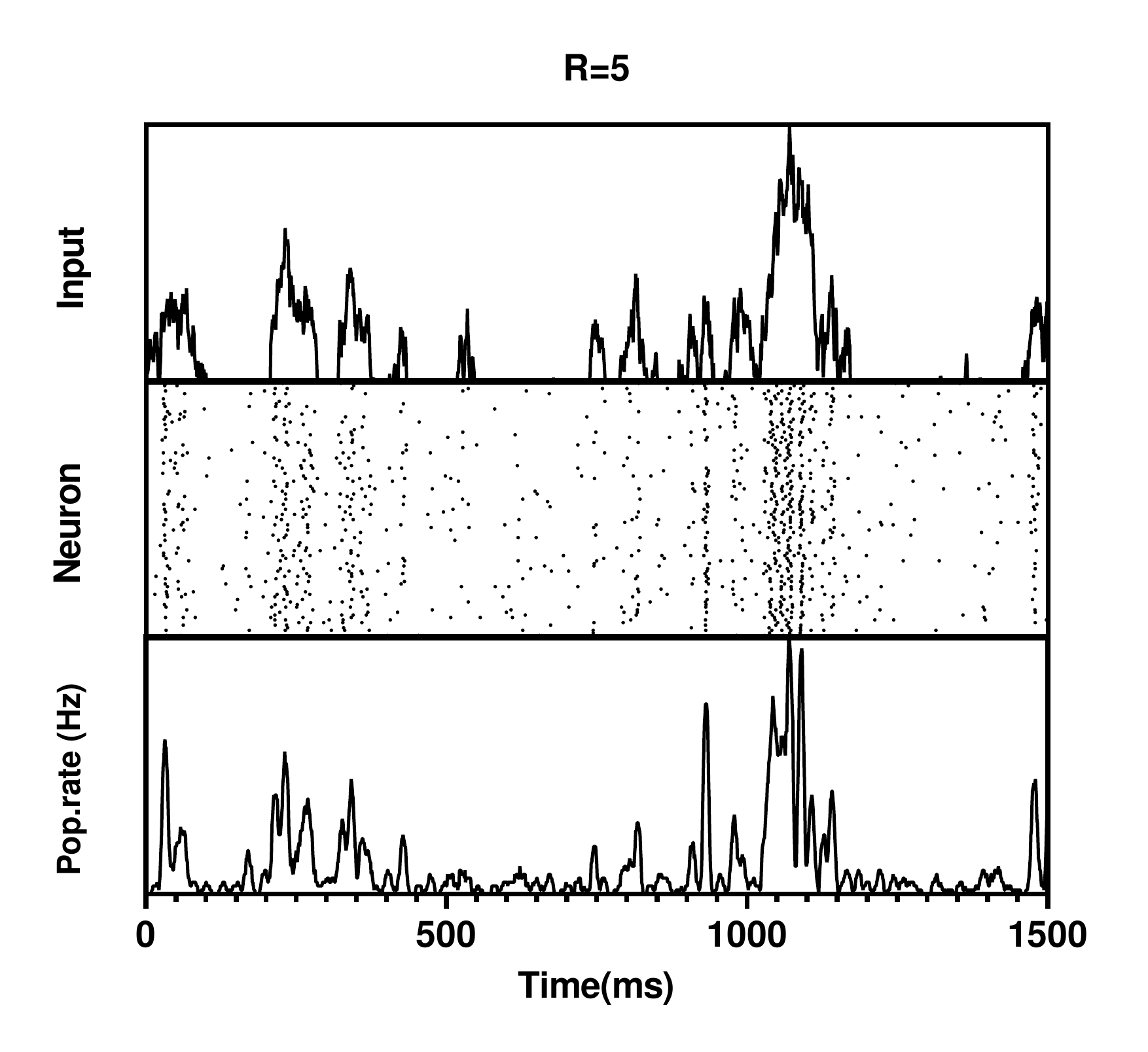}}
	\subfigure[]{
		\includegraphics[width=5.8cm]{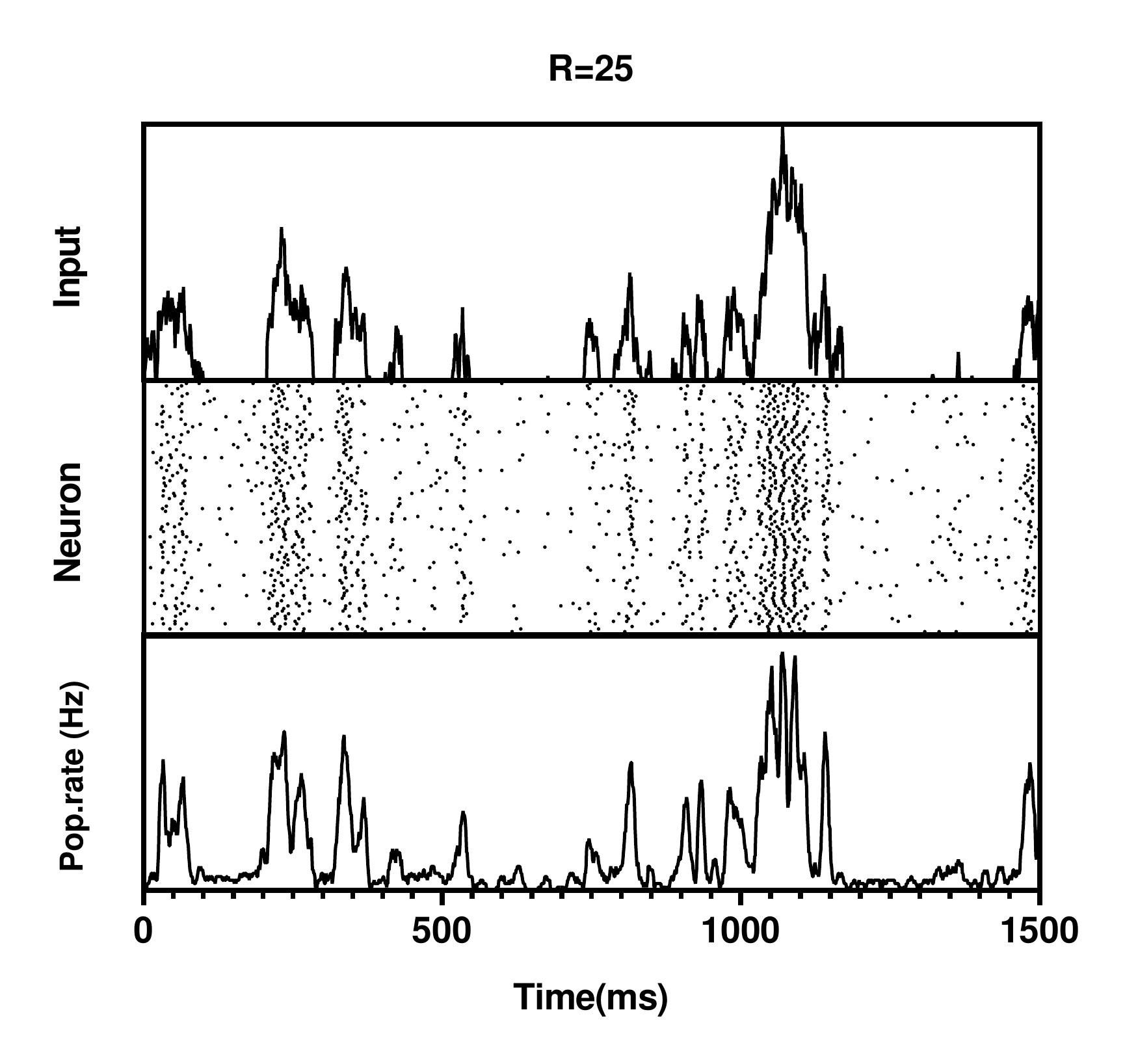}}
	\subfigure[]{
		\includegraphics[width=5.8cm]{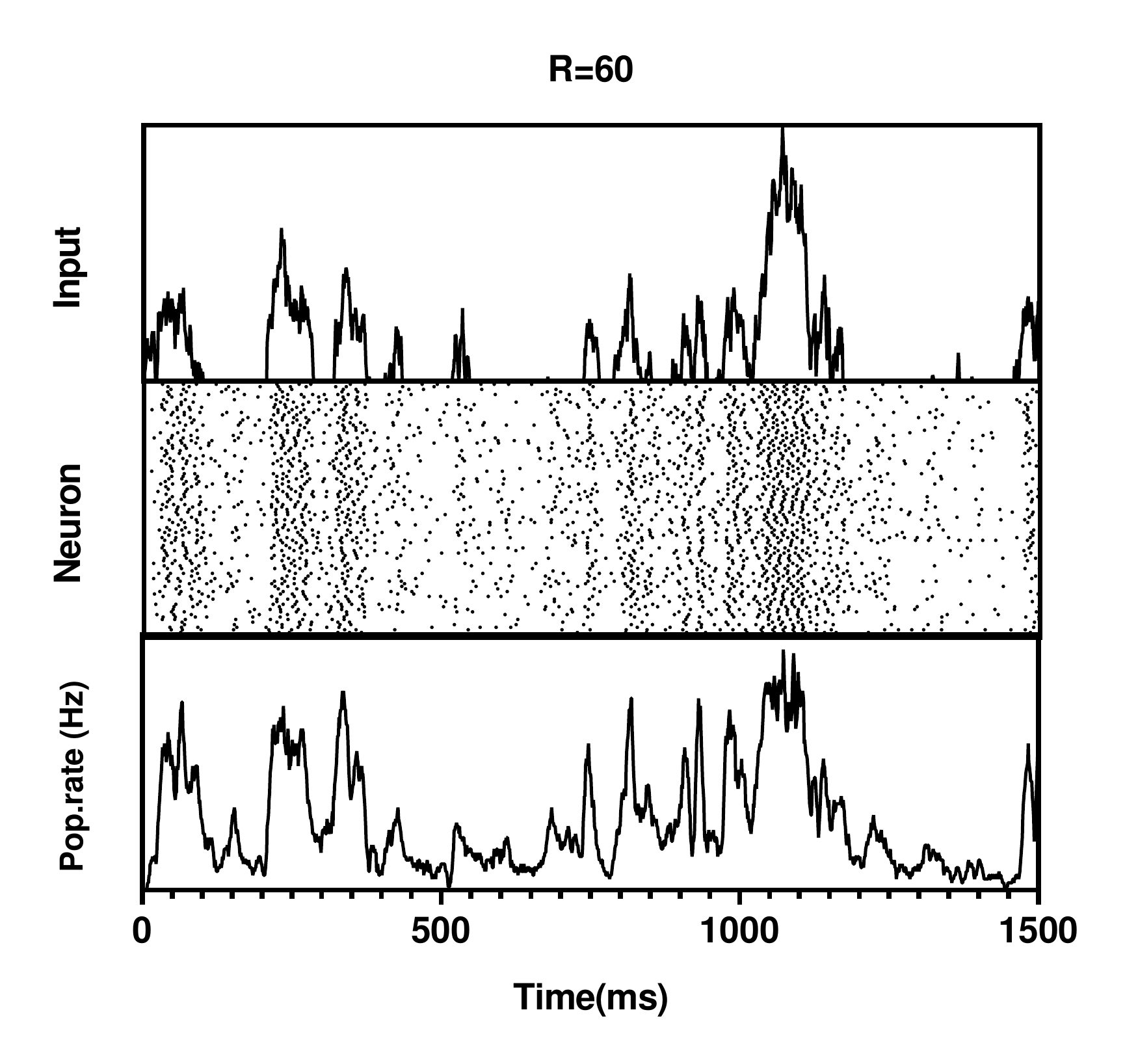}}
	\caption{(a) A glimpse that the effect of $R$ on the encoding quality $Q$. The coding quality in the first layer as a function of the I/E synaptic strength ratio $R$. The encoding quality has a better performance as $R$ locates in a moderate range. $P_{rc}=0.1$, $D=1$, $g^{exc}=0.102$, $\tau^{exc}=0.3, \tau^{inh}=0.6$. (b)(c)(d) The typical results of population firing rate encoding from the same stimuli injected into the recurrent network. From the top to the bottom are input, spike raster, and population firing rate, respectively. $(b),(c),(d)$ are $R = 5$, $R = 25$, $R = 60$, respectively, and the case of $R=25$ is of the best performance.}
	\label{fig.QvsR_raster_1}
\end{figure}

The network state has a directive effect on information encoding. How the network state affects the population rate coding quality is an important problem. In the current studied network, the ratio between inhibitory and excitatory synaptic strength directly determine the network state, making which more excitatory or inhibitory. If the ratio is small, the excitation dominates, otherwise, inhibition dominates. Meanwhile, the recurrent probability $P_{rc}$ determines how much tightly the neurons in the network connect with, which directly influences the total number of excitatory and inhibitory connections, respectively. That is to say that the I/E strength ratio $R$ and recurrent probability $P_{rc}$ definitely have a great influence on the network's state. In this subsection, therefore, we first discuss the effects of $R$ and $P_{rc}$ on the encoding quality $Q$.

We first show a curve of $Q$ versus $R$ under some certain parameters in Fig. \ref{fig.QvsR_raster_1}(a). From this glimpse that the effect of $R$ on the encoding quality $Q$, we see that there might exist an optimal range of $R$ for a better encoding performance, which in this figure suggests a range of $15<R<40$. The encoding quality $Q$ takes higher than $92\%$ denoting that the current studied network encodes the stimuli into population firing rate as much as $92\%$ or even more. It is a positive phenomenon for us because the high $Q$ suggests that the model we construct in this paper is able to encode input information and the population rate code could represent the stimuli well. The typical results are shown in the Fig. \ref{fig.QvsR_raster_1}(b)(c)(d), in which the top, middle, and bottom represent the stochastic input, network spiking raster and the corresponding population firing rate, respectively. Comparing to the other two cases, the case $R=25$ has a better performance on representing the input using population firing rate, which preserves the input more completely while in the other two cases some signals are weakened or amplified.
\begin{figure}[!htbp]
	\centering
	\includegraphics[width=8.5cm]{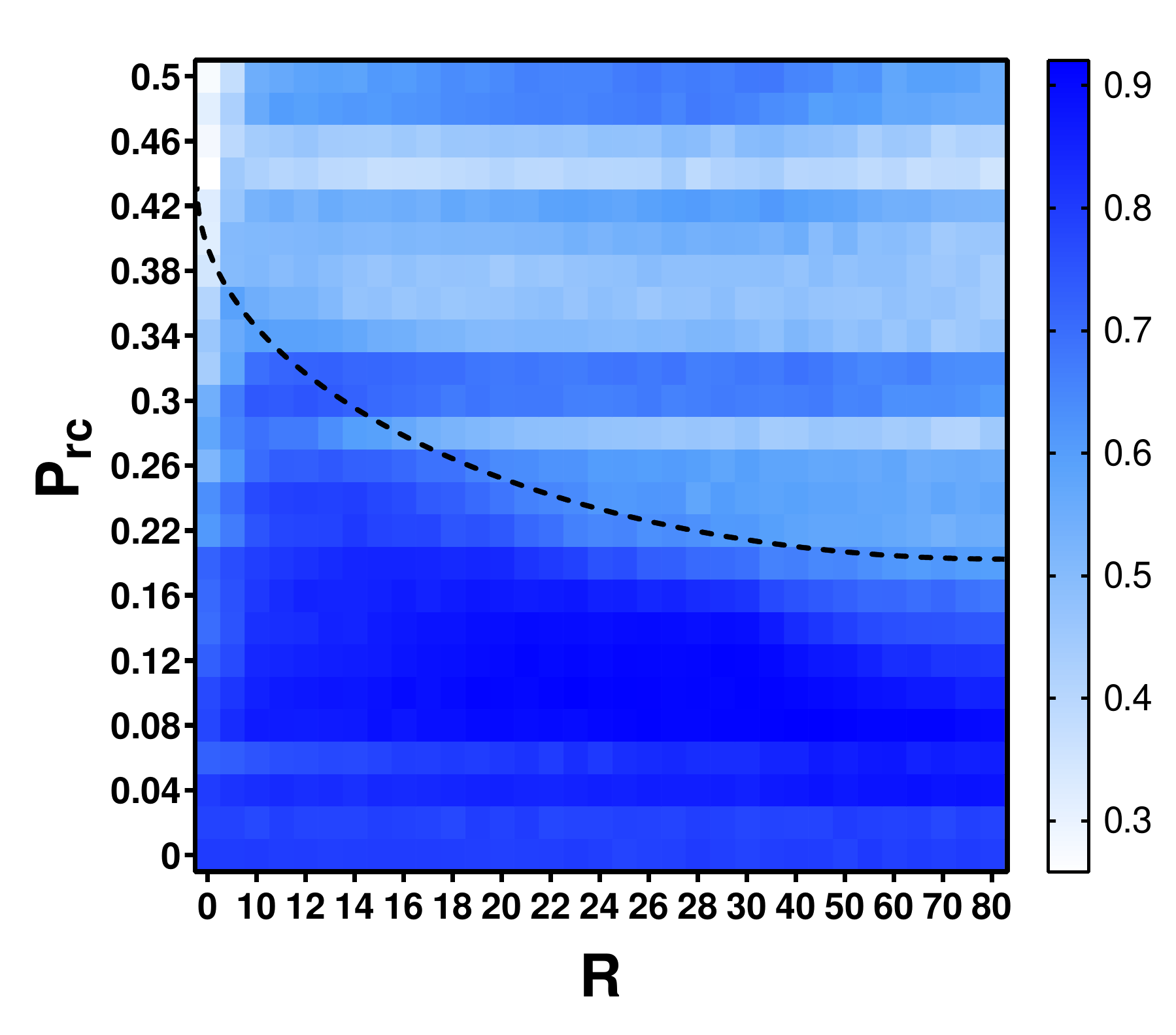}
	\caption{Encoding quality $Q$ depending on $R$ and $P_{rc}$. The color code denotes the coding quality with dark blue being the best cases. In the parameter space of $R$ and $P_{rc}$ it seems that $Q$ takes higher values below the black dashed curve(noted as blue area) while smaller values above the dashed curve(noted as white area). In the blue area, especially around $P_{rc}=0.1$, the quality $Q$ has a clear tendency to increase first and then decrease similar to the curve shown in \ref{fig.QvsR_raster_1}(a). The encoding quality has a better performance when the recurrent probability is set around 0.1. $D=1$, $g^{exc}=0.102$, $\tau^{exc}=0.3, \tau^{inh}=0.6$. (Averaged in 10 trials). }
	\label{fig.Prc_R_Q}
\end{figure}

The Fig. \ref{fig.QvsR_raster_1} obviously shows that the I/E strength ratio has an important role in the population rate coding and the firing of neurons in the network. What does it happen if $P_{rc}$ is altered? Now, we test the encoding quality versus different $R$ and $P_{rc}$. The results that effects of $Q$ on the parameters $P_{rc}$ and $R$ are shown in the Fig. \ref{fig.Prc_R_Q}. From this figure, however, we see that the curve as is shown in the Fig.\ref{fig.QvsR_raster_1}(a) is not adapted to every values of $P_{rc}$. In the parameter space of $R$ and $P_{rc}$ it seems that $Q$ takes higher values below the black dashed curve(noted as the dark blue area) while smaller values above the dashed curve(noted as the light blue area). Interestingly, for every $P_{rc}$, there exists an optimal range of $R$ in which $Q$ takes higher values while smaller values out of the range.  
\begin{figure}[!htbp]
	\centering
	\includegraphics[width=11cm]{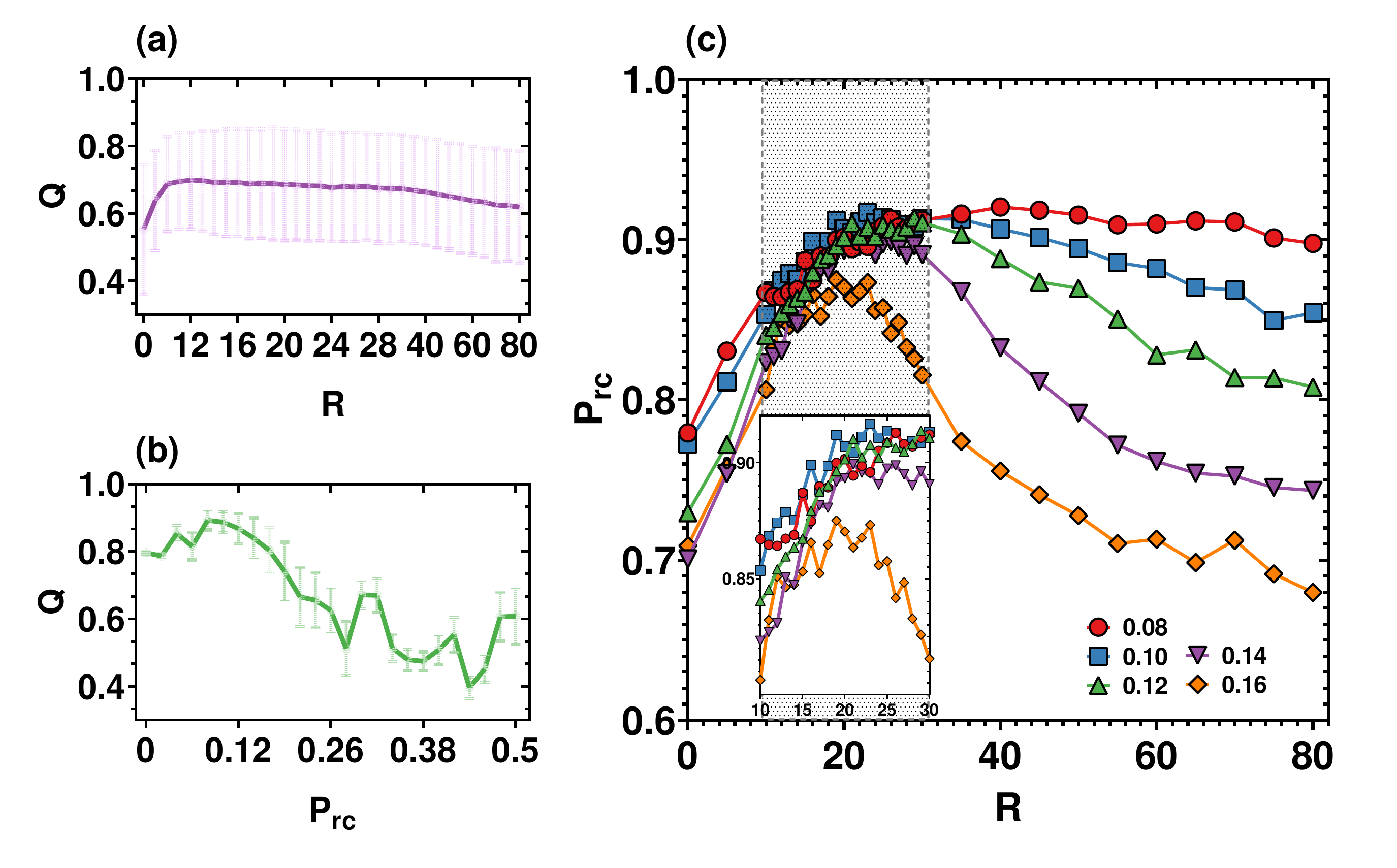}
	\caption{Statistical results and samples of data from the effects of $P_{rc}$ and $R$ on the population rate coding. (a) The purple line denotes the mean values of $Q$ vs. $R$ under different $P_{rc}$. The purple bar denotes the corresponding SD. (b) The green line denotes the mean values of $Q$ vs. $P_{rc}$ under different $R$ and the green bar denotes the corresponding SD. (c) The effects of $R$ and selected $P_{rc}$ on the population rate coding. The subgraph denotes the overlapping curves located in the gray area where $R$ takes value from $12<R<30$. The encoding quality performs worse in the case of low and high values of $R$ while better in the moderate values of $R$. $0.08 <= P_{rc} <=0.16$. }
	\label{fig.stasitics_sample_Prcs_R_Q}	
\end{figure}

From the statistical characteristic shown in the Fig.  \ref{fig.stasitics_sample_Prcs_R_Q}(a), the mean values of $Q$, however, present a decaying trend as $R$ increases, in which the large error bars are almost due to the lower encoding quality in the light blue area. Therefore, we should focus on the dark blue area with high encoding performance. The green curve in Fig.  \ref{fig.stasitics_sample_Prcs_R_Q}(b) shows that $Q$ presents a decaying trend with increasing $P_{rc}$ and takes optimal values around $P_{rc}\approx0.1$. According to the two statistical results and corresponding analysis, we now turn to the Fig. \ref{fig.Prc_R_Q} again.
Obviously, most optimal ranges of $R$ located in the dark blue area, though the range gets narrower as $P_{rc}$ increases. In the dark blue area, especially around $P_{rc}=0.1$, the quality $Q$ has a clear tendency to increase first and then decrease similar to the curve shown in \ref{fig.QvsR_raster_1}(a). We call this curve with such a tendency the idealized curve. The encoding quality has a better performance as the recurrent probability is set around 0.1. 

In order to determine the relationship of $Q$ versus $P_{rc}$ around $R\approx0.1$, we show some curves selected from the results around $R\approx0.1$ in the Fig. \ref{fig.Prc_R_Q}. As the Fig. \ref{fig.stasitics_sample_Prcs_R_Q} illustrates, almost all of the five curves have the idealized tendency of $Q$ versus $R$. However, in details, the curve tends to be saturated as $R$ increases if $P_{rc}>0.12$; the curve tends to decay rapidly as $R$ increases if $P_{rc}<0.12$. In other word, the optimal range tends to be wider for $P_{rc}>0.12$ while narrower for $P_{rc}<0.12$. In the following simulations, we choose $P_{rc}=0.12$ cause that it is a compromise of the previous two cases which has no obvious saturation and sharp decay but a suitable optimal range of $R$.
 
\subsection{Effects of considered parameters on population rate coding}
\subsubsection{Noise intensity}
\begin{figure}[!htbp]
	\centering
	\includegraphics[width=9 cm]{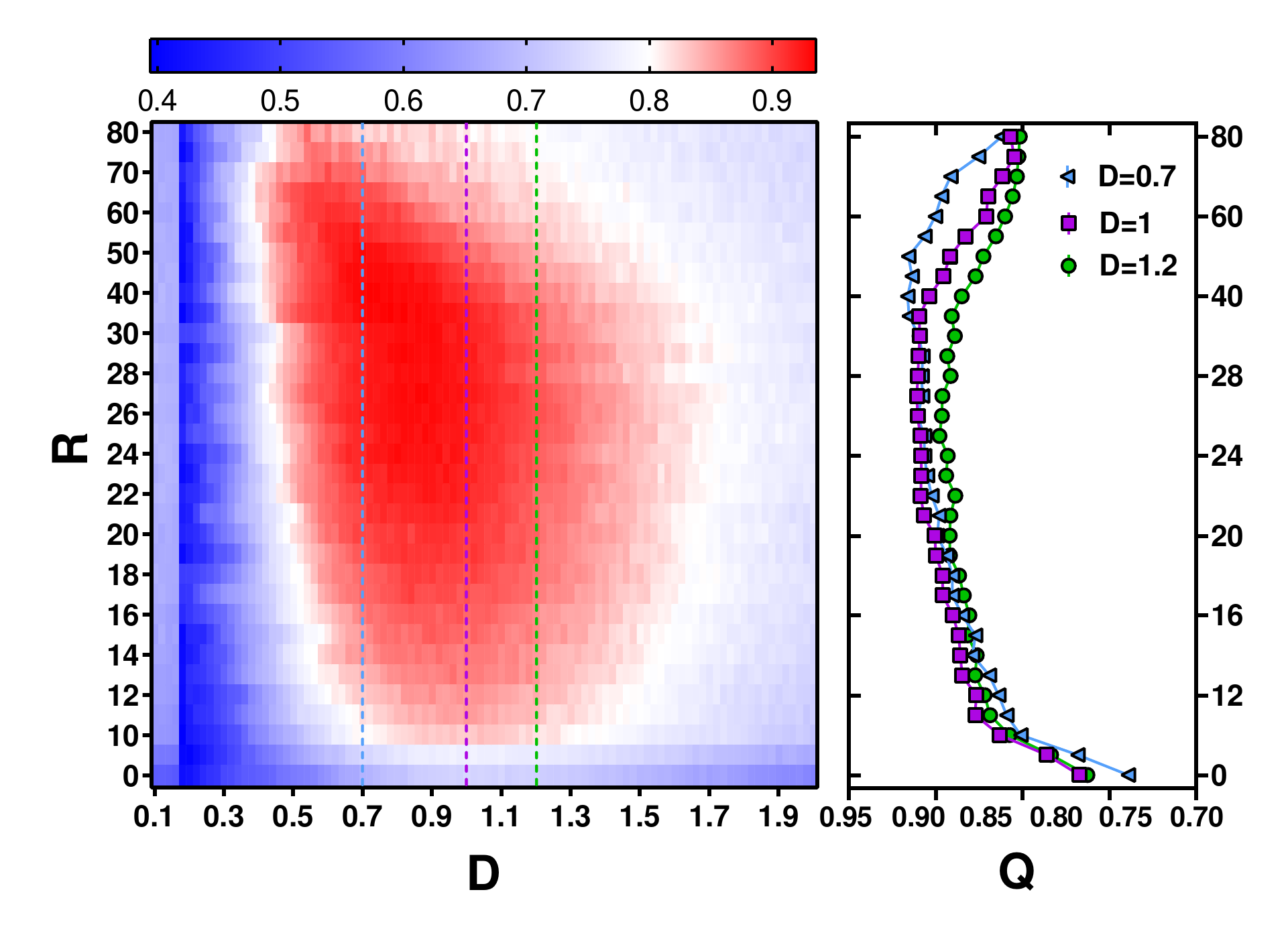}
	\caption{Encoding quality depending on $D$ and $R$. The color denotes the encoding quality and red denotes higher $Q$. We slice three cases of $D=0.7,1,1.2$ in the left part with different color and plot the corresponding curve of $Q$ vs. $R$ in the right part. $P_{rc}=0.12$, $g^{exc}=0.102$, $\tau^{exc}=0.3, \tau^{inh}=0.6$.}
	\label{fig.QvsR_noise}	
\end{figure}

Neurons in the biological environment receive background noise from time to time and fire spikes spontaneously. In the studied work, we use Gaussian noise to mimic the background noise and spontaneous spikes. It turned out that, in the previous studies, the noise intensity is a key parameter for information coding\cite{Rossum2002,Wang2006}. The encoding quality depending on the noise intensity $D$ vs. $R$ is shown in the Fig. \ref{fig.QvsR_noise}.

\begin{figure}[!htbp]
	\centering
	\includegraphics[width=11 cm]{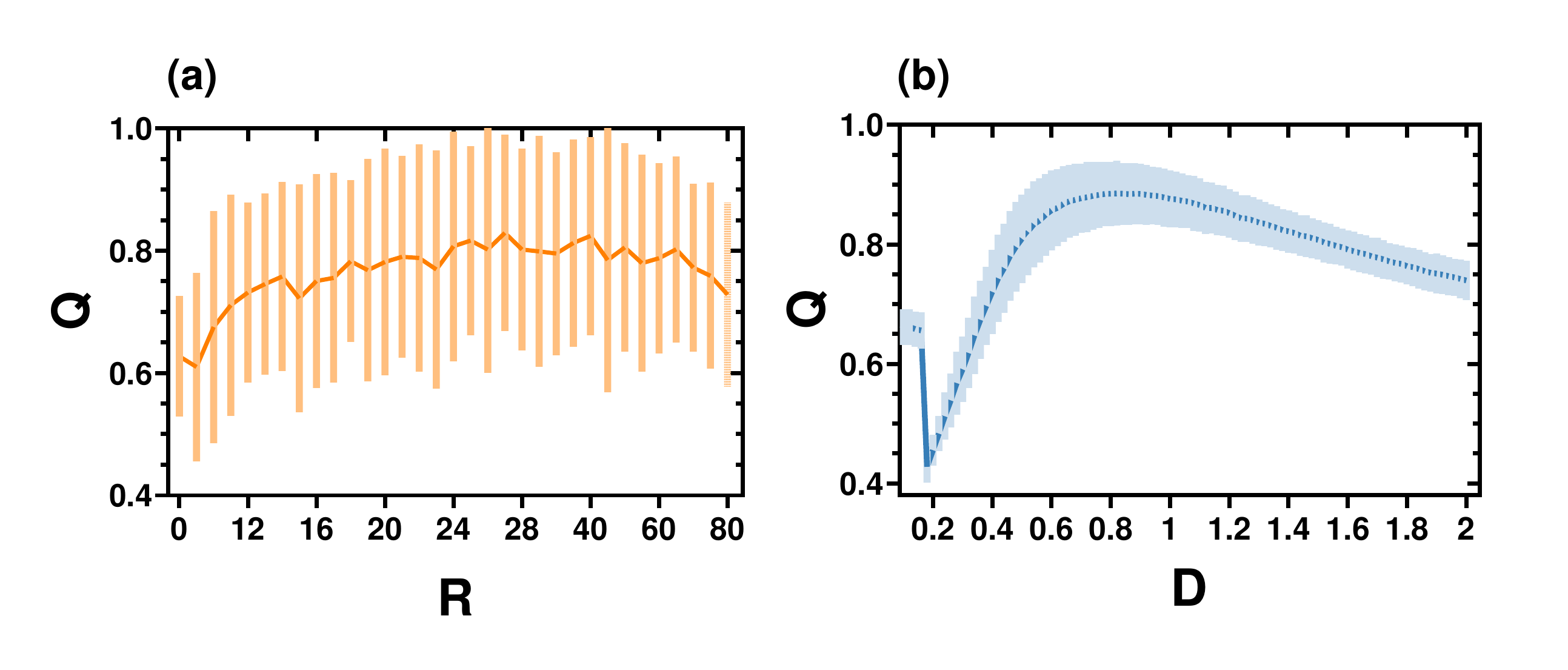}
	\caption{Statistics of the data from the effects of noise intensity on the population rate coding. (a) The orange line denotes the mean values of $Q$ vs. $R$ under different $D$. The orange bar denotes the corresponding SD. The curve shows the same trends that lower $Q$ for lower $R$ and higher $Q$ for moderate $R$.(b) The blue line denotes the mean values of $Q$ vs. $D$ under different $R$ and the blue bar denotes the corresponding SD. With the exception of the cases $D<0.2$, in most values of $D$ the encoding quality increases as $D$ increases and decay for the large $D$, which means there exist an optimal range of $D$ for high encoding quality.}
	\label{fig.statistics_noise}	
\end{figure}

As is shown in the Fig. \ref{fig.QvsR_noise}, the left part denotes the encoding quality $Q$ on the parameter space of $R$ and noise intensity $D$. The red color represents better performance on information coding. Obviously there exist an optimal area of $R$ and $D$ in which $Q$ takes high value(red area). For $0.6<D<1.2$, the network encodes the input information very well within an optimal range of $R$. Meanwhile, in the range of $0.6<D<1.2$, the lower limits and upper limits of the optimal range of $R$(eg., the case that $Q>0.9$) changes with varying $D$. In details, we found that as $D$ is small the upper and lower limits of $R$ are both larger, otherwise they are smaller. To see the relationship between $Q$ and $D, R$, we perform some statistical analysis of the data from the effects of noise intensity on the population rate coding. The statistical results are shown in the Fig. \ref{fig.statistics_noise}. 
The orange curve shows the same trend that lower $Q$ for lower $R$ and higher $Q$ for moderate $R$. Notably, the trend is weaker than that we 'see' from the Fig. \ref{fig.QvsR_noise} in vision due to the bad performance out of the red area, but that does not affect us to get such a conclusion. Similarly, the blue curve in the Fig. \ref{fig.QvsR_noise}(b) tells us that there exists an optimal range of $D$ where the encoding quality takes high values.

Take a closer look at the results.  We slice three cases of $D=0.7,1,1.2$ in the left part of Fig.  \ref{fig.QvsR_noise} with different color and plot the corresponding curve of $Q$ vs. $R$ in the right part. As we discussed above, the three cases all have an optimal range of $R$, the case $D=0.7$ has a wider optimal range of $R$ and its lower and upper limits are larger, the case $D=1.2$ has a narrower optimal range of $R$ and its lower and upper limits are smaller. Comparing to the other cases, the case $D=1$ has a moderate optimal range of $R$  maintaining the high level of $Q$ and $Q$ does not decay rapidly if $R$ exceeds the corresponding optimal range, which makes the case $D=1$ be a compromise of the other two cases and proves the rationality of the parameter $D=1$ that we selected before. Of course, we could get some very high encoding quality with some noise intensity like $D=0.7$. We still select $D=1$ in the later simulations because that does not affect the results we investigate the effects of other parameters on the encoding performance.

\subsubsection{Excitatory synaptic strength}
\begin{figure}[!htbp]
	\centering
	\includegraphics[width=7.5 cm]{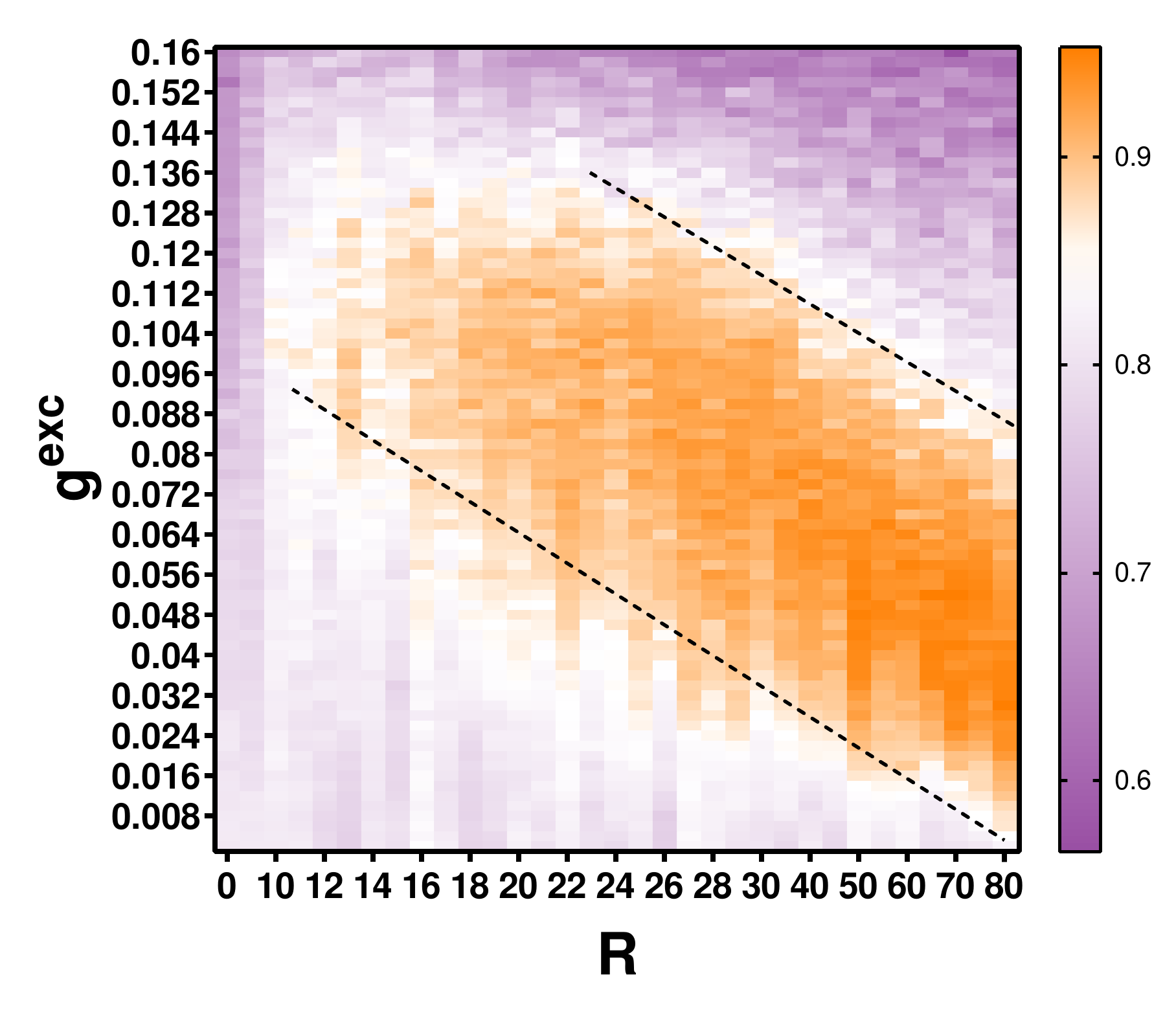}
	\caption{Encoding quality depending on $g^{exc}$ and $R$. The color decodes the encoding quality and the orange color denotes higher $Q$. $Q$ takes higher values in the area between the two black dashed parallel lines but lower values out of the area especially in the upper right corner. The underlying mechanism might is the large $R \dot g^{exc}$ namely $g^{inh}$ depresses the synaptic activity in the network making the encoding quality worse. $P_{rc}=0.12$, $D=1$, $\tau^{exc}=0.3$, $\tau^{inh}=0.6$.}
	\label{fig.QvsR_gexc}	
\end{figure}

The excitatory synaptic strength is a key parameter who directly determines the strength of excitatory synapses and indirectly determines the inhibitory synaptic strength together with $R$. Therefore, the excitatory synaptic strength $g^{exc}$ together with I/E ratio $R$ have a great influence on the network state and the performance of population rate coding. The encoding quality depending on the $g^{exc}$ and $R$ is shown in the Fig. \ref{fig.QvsR_gexc}.

As is shown in the figure, there is an obvious orange area denoting the better performance where most of the encoding quality is almost beyond 0.9, which is located between the black dashed parallel lines. We note that values of $Q$ is high in the orange area almost overlapping with the line whose slope approximates $-1$ from the lower-right corner to the upper-left corner, while $Q$  outside the orange area is lower. To describe the graph in details, $Q$ takes higher values in the middle area on the 2-D parameter space(orange color), second-high values in the lower-left corner(white color) and low values in the upper-right corner(purple color). Meanwhile, we note that the product of $R$ and $g^{exc}$ takes lower values in the lower left corner, moderate values in the middle area and high values in the upper right corner. It seems that the product of $R$ and $g^{exc}$ has a positive relationship with the encoding quality. The product of $R$ and $g^{exc}$ is exactly the inhibitory synaptic strength $g^{inh}$. The inhibitory synaptic strength might have a great influence on the encoding quality, which is understood as follows. For a large $g^{inh}$, the encoding quality is worse, which suggests that the too strong inhibitory synapses depress the encoding ability of the network making $Q$ smaller; for a small  $g^{inh}$, $Q$ is larger than the values for large $g^{inh}$ but still smaller than $Q$ for intermediate $g^{inh}$, which suggests that the decreasing inhibitory strength could facilitate the encoding quality of the network but too weak inhibition could indirectly excite the network activity to decrease the accuracy of information coding. For intermediate $g^{inh}$, the network could keep in an intermediate state — not too excitatory or too inhibitory — so that the network could respond to the signals well and possess a better performance for coding. Notably, it explains why in the results that we fix the value of $g^{exc}$ above there exists an optimal range of $R$, namely, the network needs an intermediate state to have a better coding ability. 

In a summary, we find the effects of $g^{inh}$ on the encoding quality, which is determined by the excitatory synaptic strength $g^{exc}$ and I/E ratio $R$. That's to say the excitatory synaptic strength and I/E ratio influence the encoding quality together. For a $g^{exc}$(not too large), there always exists an optimal range of $R$ locating in the intermediate values of $R$. We, therefore, fix the value of $g^{exc}$, e.g., $g^{exc}=0.102$, in the previous simulations which is enough for our works. 
\begin{figure}[!htbp]
	\centering
	\includegraphics[width=7.5 cm]{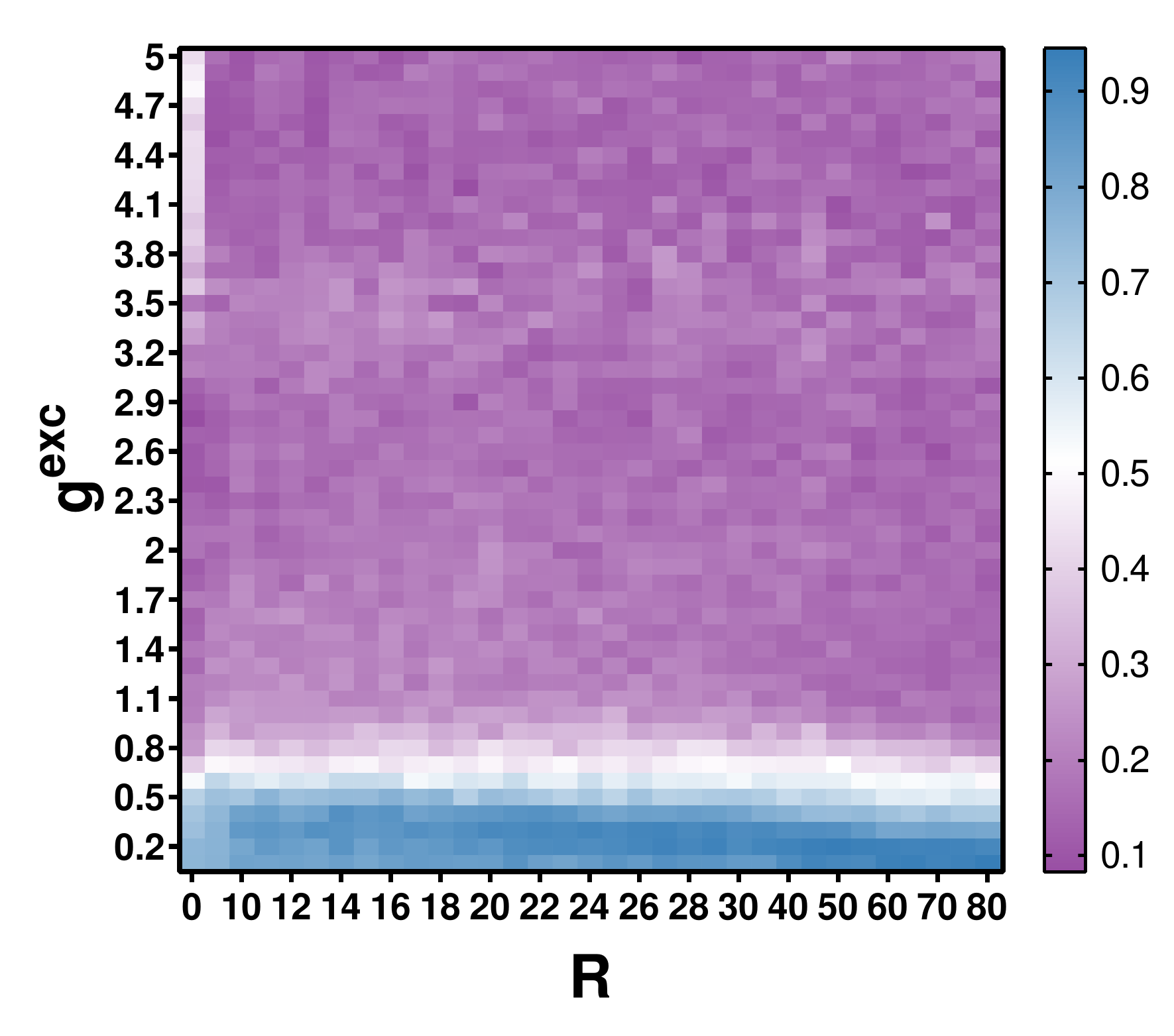}
	\caption{The effects of synaptic time constant on the population rate coding. The color decodes the encoding quality and the blue color denotes higher $Q$. $P_{rc}=0.12$, $D=1$, $g^{exc}=0.102$.}
	\label{fig.QvsR_tau}	
\end{figure}

\subsubsection{Synaptic time constant}
In this subsection, we investigate the effects of synaptic time constant on the encoding performance. In the simulations, we fix the ratio between the excitatory and inhibitory synaptic time constant $\tau^{inh} = 2 \,\tau^{exc} $, then we alter the values of $\tau^{exc}$. The results are shown in the Fig. \ref{fig.QvsR_tau}, which suggests that only small synaptic time constant is good for encoding performance. It is easy to understand that only if a synapse with a small synaptic time constant, namely, fast dynamics, could respond to the coming synaptic currents with high speed so that the coding is more accurate.

\subsection{Comparison with the determined network}
We have studied the effects of multiple parameters on the population rate coding quality of the neuronal network, in which we claimed that one presynaptic neuron might perform either excitatory or inhibitory effects on the corresponding postsynaptic neurons. That is different from the previous works in which they considered the neurons into determined types — inhibitory or excitatory. In order to determine the differences, we compare the population rate coding quality between the previous networks(determined EI model) and our studied network(undetermined EI model). The synaptic strength and noise intensity are mainly considered in this subsection.

First, let us introduce the set up of the determined EI model. The network consists of 100 neurons with 80 excitatory neurons and 20 inhibitory neurons, which could be named as excitatory subpopulation $E$ and inhibitory subpopulation $I$. The connection probability of these subpopulations is $P_{EE}=P_{EI}=P_{IE}=P_{II}=0.12$ denoting excitatory-excitatory connection, excitatory-inhibitory connection, inhibitory-excitatory connection, and inhibitory-inhibitory connection, respectively. The other parameters are the same as the parameters in our considered network mentioned above.
\begin{figure}[!htbp]
	\centering
	\includegraphics[width=11.5 cm]{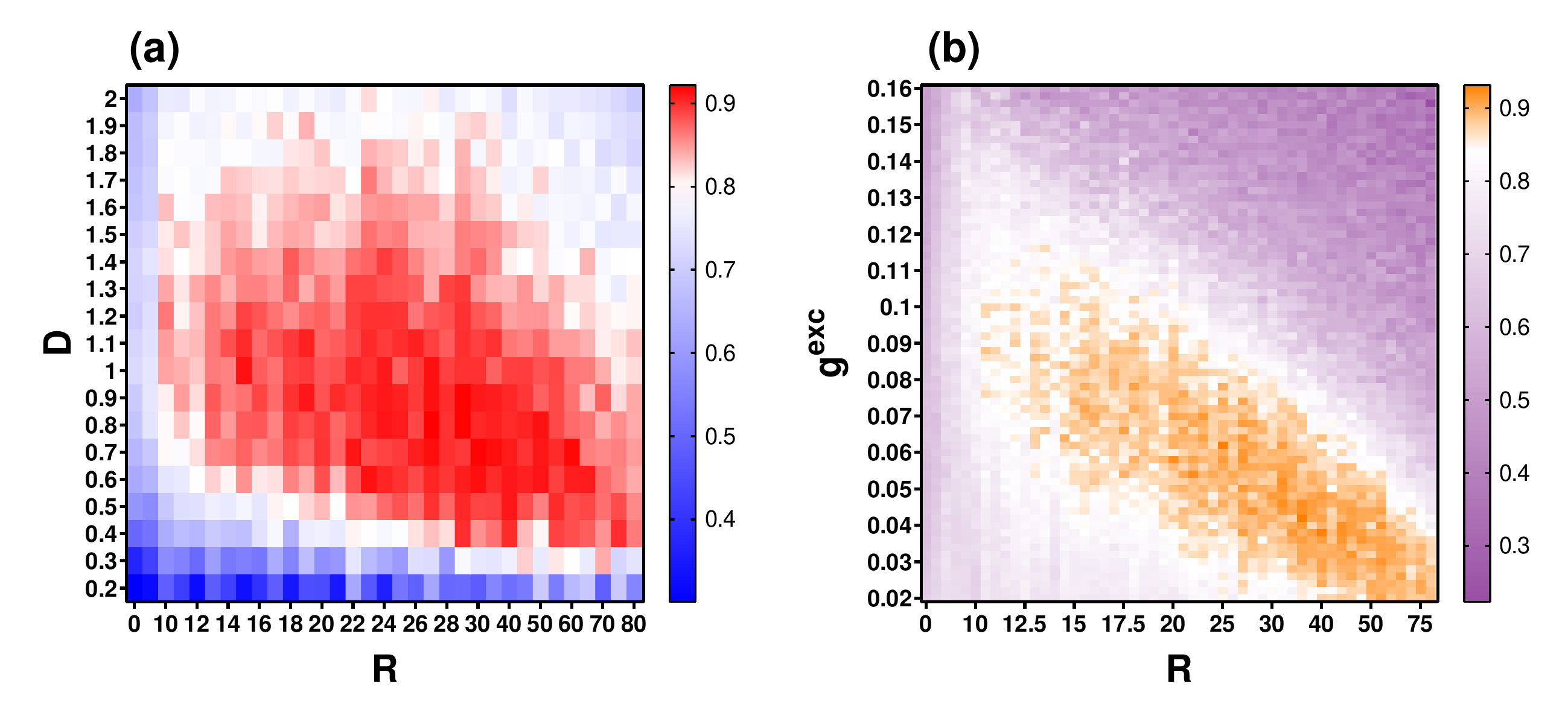}
	\caption{Effects of synaptic strength and noise intensity on the encoding quality in determined EI model. (a) The noise intensity has a similar effect with us on the encoding quality that too weak and strong noise depress the population rate coding while intermediate noise facilitate coding. $g^{exc}=0.08$. (b) The synaptic strength has a similar effect with us on the encoding quality that only intermediate $R \dot g^{exc}$ has the optimal effect on the population rate coding. $D=0.7$.}
	\label{fig.QvsR_comparison_g_D}	
\end{figure}

\begin{figure}[!htbp]
	\centering
	\includegraphics[width=12.5 cm]{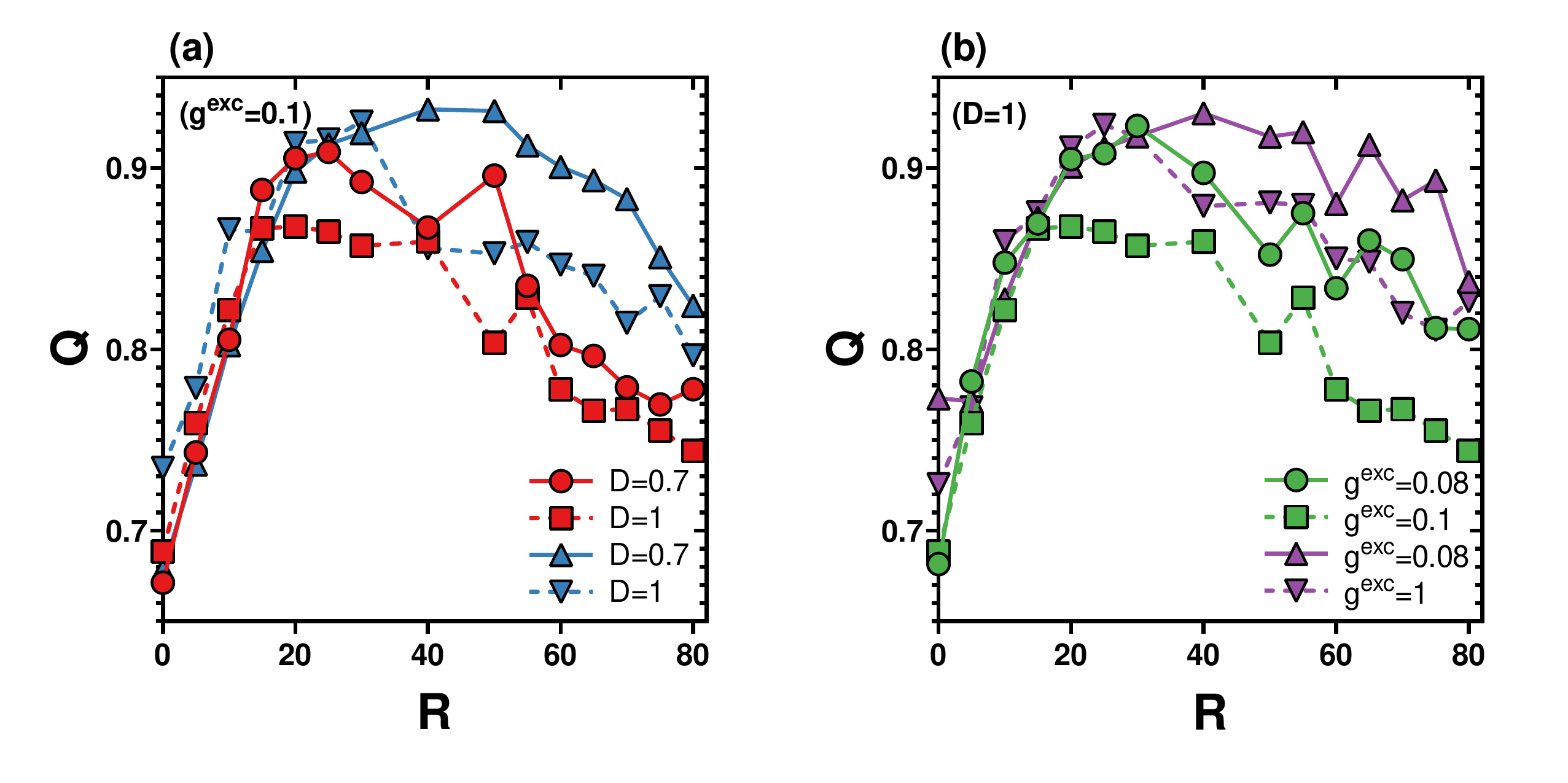}
	\caption{Comparison of synaptic strength and noise intensity on the encoding quality in two models. (a) The red color denotes the determined model while blue color denotes our model. In both models, as $R$ increases, $Q$ reaches the optimal value sooner under large noise intensity than the counterpart under the smaller noise intensity, but its value is smaller. The optimal $Q$ takes higher values in our model than the counterpart in the determined model. $g^{exc}=0.1$. (b) The green color denotes the determined model while the purple color denotes our model. In both models, as $R$ increases, $Q$ reaches the optimal value sooner under large synaptic strength than the counterpart under the smaller synaptic strength, but its value is smaller. The optimal $Q$ takes higher values in our model than the counterpart in the determined model. $D=0.7$.}
	\label{fig.QvsR_comparison_g_D1}	
\end{figure}

The results of the determined model are illustrated as in the Fig.  \ref{fig.QvsR_comparison_g_D}. As is shown in the subgraph(a), the noise intensity has a similar effect with ours on the encoding quality that too weak and strong noise depress the population rate coding while intermediate noise facilitates coding. The synaptic strength has a similar effect with us on the encoding quality that only intermediate $R * g^{exc}$ has the optimal effect on the population rate coding. We note that the values of $g^{exc}$ and $R$ both are smaller than the counterparts in our model, which might be due to the determined excitatory and inhibitory neurons have the different and separating effects so that they do not need some larger parameters. With the same parameters, however, the encoding quality in the determined model seems worse than ours like the Fig. \ref{fig.QvsR_comparison_g_D}(b). To confirm that point, we show comparisons of some cases in the following figures. The comparison of noise intensity on the encoding quality in two models is shown in Fig. \ref{fig.QvsR_comparison_g_D1}(a), which shows us that in both models, as $R$ increases, $Q$ reaches the optimal value sooner under large noise intensity than the counterpart under the smaller noise intensity, but its value is smaller. The optimal $Q$ takes higher values in our model than the counterpart in the determined model. As is shown in the Fig. \ref{fig.QvsR_comparison_g_D1}(b), the synaptic strength has a similar effect and better performance in our model than the determined model.
\section{Conclusion}

Information coding in cortical network is one of critical questions for people to understand the brain, which has attracted extensive attention. In the past decades, people developed many potential strategies of information coding\cite{Newsome1989,Rossum2002,Diesmann1999,Vogels2005,Thorpe1996,Thorpe2001,Rullen2001,Gollisch2008,Olshausen2004}, in which there are two main types of coding paradigm including temporal coding\cite{Aertsen1996,Diesmann1999,Litvak2003} and rate coding\cite{Newsome1989,Rossum2002}. Among thses neural code hypthesis, the population rate coding has been widely studied in many works\cite{Brunel1999,Knight2000,Gerstner2000,Dayan2005,Masuda2003DualityOR,Masuda2005,Rossum2002,Wang2009,Guo2012}.

People constructed many recurrent networks consisting of pre-determined excitatory and inhibitory neurons to model the cortical neuronal networks. In these works, whether the synaptic connections are excitatory or inhibitory is mainly determined by the type of presynaptic neurons. Considering the physiological evidences, however, the property of the synaptic connection should be determined by the type of activated receptors. Many experimental works illustrated that both excitatory and inhibitory receptors might co-exist and in the same presynaptic synapse\cite{Shrivastava2011,Kantamneni2015}, as well as the cross-talk between excitatory and inhibitory receptors plays a key role in the balance of excitation and inhibition in brain\cite{Kantamneni2015}. Inspired by these evidences, therefore, we construct a recurrent neuronal network in which one presynaptic neuron might perform excitatory or inhibitory effect on the corresponding postsynaptic neurons.

In this paper, we study the population rate coding in our considered neuronal network. The I/E strength ratio $R$ is a key parameter that determines how different between the excitatory and inhibitory synaptic strength, which indirectly influences the network state. We find there exists an optimal range of $R$ for better population rate coding performance, which usually locates in the intermediate values of $R$. The recurrent probability has a similar optimal intermediate range in which the population rate coding performs well. After determining the two key parameters, in the following simulations effects of the noise intensity, synaptic strength and synaptic time constant on the population rate coding are investigated. The noise intensity has an obvious effect on the population rate coding that intermediate noise intensity could facilitate the encoding quality while large and small noise intensity depresses the encoding quality. Excitatory synaptic strength together with I/E strength ratio influences the encoding quality illustrating that if the excitatory synaptic strength is not too large there exists an optimal range of $R$. The optimal range of $R$ slides with the increasing excitatory synaptic strength. The synaptic time constant determines the speed of neurons' response to signals so that small synaptic time constant promotes the accuracy of population rate coding. All the results above suggest that the optimal range of every parameters(besides synaptic time constant) locates in the intermediate values. With the suitable combination of these parameters the neuronal network will encode signal information in population rate code very well.

Furthermore, we compare the population rate coding performance of our considered network(undetermined EI model) and the previous networks(determined EI model). We find that the determined EI model has a similar optimal range of $R$, but $R$ in the model usually takes lower than ours. The noise intensity also has a similar effect on the population rate coding that too weak and too strong noise both are bad for the population rate coding. The synaptic strength has a little different effect on the encoding quality. Weak excitatory synaptic strength could enhance the performance of population rate coding while the strong counterpart decreases the performance. Notably, with the same setting parameters, the population rate coding performance in the determined model is worse than ours. In conclusions, our simulation results suggested that the neuronal network with determined types of neurons has a similar effect as the network without determined types of neurons(ours) on the performance of population rate coding, but our model has a better performance in population rate coding and a more rational architecture to some degree.

Although we improve the neuronal network by considering the actual effect of synaptic connections determined by the types of activated receptors, it is still not very biological enough. In fact, many unknown issues about the co-release of excitatory and inhibitory synaptic receptors remain. There are lots of further works on this topic. On the one hand, the more complex and realistic synapse model should be considered such as introducing the stochastic process of neural transmitter and receptor. On the other hand, apart from the population rate coding, it seems there are kinds of more accurate code representing the neural information, which are worth studying to open our minds.

\section*{Acknowledgements}
		This work is supported by the National Natural Science Foundation of China (Grant Nos. 11472061, 11572084) and the Fundamental Research Funds for the Central University (No. 2018XKJC02).

	\bibliography{Popratecoding}	

\begin{thebibliography}{10}
\expandafter\ifx\csname url\endcsname\relax
  \def\url#1{\texttt{#1}}\fi
\expandafter\ifx\csname urlprefix\endcsname\relax\def\urlprefix{URL }\fi
\expandafter\ifx\csname href\endcsname\relax
  \def\href#1#2{#2} \def\path#1{#1}\fi

\bibitem{deCharms2000}
R.~C. deCharms, A.~Zador, Neural representation and the cortical code, Annual
  Review of Neuroscience 23~(1) (2000) 613--647, pMID: 10845077.
\newblock \href {http://dx.doi.org/10.1146/annurev.neuro.23.1.613}
  {\path{doi:10.1146/annurev.neuro.23.1.613}}.

\bibitem{Newsome1989}
W.~T. Newsome, K.~H. Britten, J.~A. Movshon, Neuronal correlates of a
  perceptual decision, Nature 341 (1989) 52--54.
\newblock \href {http://dx.doi.org/10.1038/341052a0}
  {\path{doi:10.1038/341052a0}}.

\bibitem{Georgopoulos1993}
A.~Georgopoulos, M.~Taira, A.~Lukashin, Cognitive neurophysiology of the motor
  cortex, Science 260~(5104) (1993) 47--52.
\newblock \href {http://dx.doi.org/10.1126/science.8465199}
  {\path{doi:10.1126/science.8465199}}.

\bibitem{Shadlen1994}
M.~N. Shadlen, W.~T. Newsome, Noise, neural codes and cortical organization,
  Current Opinion in Neurobiology 4~(4) (1994) 569 -- 579.
\newblock \href
  {http://dx.doi.org/https://doi.org/10.1016/0959-4388(94)90059-0}
  {\path{doi:https://doi.org/10.1016/0959-4388(94)90059-0}}.

\bibitem{Marsalek1997}
P.~Marsalek, C.~Koch, J.~H.~R. Maunsell, On the relationship between synaptic
  input and spike output jitter in individual neurons., Proceedings of the
  National Academy of Sciences of the United States of America 94 2 (1997)
  735--40.
\newblock \href {http://dx.doi.org/10.1073/pnas.94.2.735}
  {\path{doi:10.1073/pnas.94.2.735}}.

\bibitem{Rossum2002}
M.~C.~W. van Rossum, G.~G. Turrigiano, S.~B. Nelson, Fast propagation of firing
  rates through layered networks of noisy neurons, Journal of Neuroscience
  22~(5) (2002) 1956--1966.
\newblock \href {http://dx.doi.org/10.1523/JNEUROSCI.22-05-01956.2002}
  {\path{doi:10.1523/JNEUROSCI.22-05-01956.2002}}.

\bibitem{Mazurek2002}
M.~E. Mazurek, M.~N. Shadlen, Limits to the temporal fidelity of cortical spike
  rate signals, Nature Neuroscience 5 (2002) 463--471.
\newblock \href {http://dx.doi.org/10.1038/nn836} {\path{doi:10.1038/nn836}}.

\bibitem{Aertsen1996}
A.~Aertsen, M.~Diesmann, M.~Gewaltig, Propagation of synchronous spiking
  activity in feedforward neural networks, Journal of Physiology-Paris 90~(3)
  (1996) 243 -- 247.
\newblock \href
  {http://dx.doi.org/https://doi.org/10.1016/S0928-4257(97)81432-5}
  {\path{doi:https://doi.org/10.1016/S0928-4257(97)81432-5}}.

\bibitem{Riehle1997}
A.~Riehle, S.~Gr{\"u}n, M.~Diesmann, A.~Aertsen, Spike synchronization and rate
  modulation differentially involved in motor cortical function, Science
  278~(5345) (1997) 1950--1953.
\newblock \href {http://dx.doi.org/10.1126/science.278.5345.1950}
  {\path{doi:10.1126/science.278.5345.1950}}.

\bibitem{Diesmann1999}
M.~Diesmann, M.-O. Gewaltig, A.~Aertsen, Stable propagation of synchronous
  spiking in cortical neural networks, Nature 402 (1999) 529--533.
\newblock \href {http://dx.doi.org/10.1038/990101} {\path{doi:10.1038/990101}}.

\bibitem{Litvak2003}
V.~Litvak, H.~Sompolinsky, I.~Segev, M.~Abeles, On the transmission of rate
  code in long feedforward networks with excitatory{\textendash}inhibitory
  balance, Journal of Neuroscience 23~(7) (2003) 3006--3015.
\newblock \href {http://dx.doi.org/10.1523/JNEUROSCI.23-07-03006.2003}
  {\path{doi:10.1523/JNEUROSCI.23-07-03006.2003}}.

\bibitem{Vogels2005}
T.~P. Vogels, L.~F. Abbott, Signal propagation and logic gating in networks of
  integrate-and-fire neurons, Journal of Neuroscience 25~(46) (2005)
  10786--10795.
\newblock \href {http://dx.doi.org/10.1523/JNEUROSCI.3508-05.2005}
  {\path{doi:10.1523/JNEUROSCI.3508-05.2005}}.

\bibitem{Masuda2002}
N.~Masuda, K.~Aihar, Bridging rate coding and temporal spike coding by effect
  of noise, Phys. Rev. Lett. 88 (2002) 248101.
\newblock \href {http://dx.doi.org/10.1103/PhysRevLett.88.248101}
  {\path{doi:10.1103/PhysRevLett.88.248101}}.

\bibitem{Masuda2003DualityOR}
N.~Masuda, K.~Aihara, Duality of rate coding and temporal coding in
  multilayered feedforward networks, Neural Computation 15 (2003) 103--125.
\newblock \href {http://dx.doi.org/10.1162/089976603321043711}
  {\path{doi:10.1162/089976603321043711}}.

\bibitem{Masuda2004}
N.~Masuda, K.~Aihara, Dual coding and effects of global feedback in
  multilayered neural networks, Neurocomputing 58-60 (2004) 33 -- 39,
  computational Neuroscience: Trends in Research 2004.
\newblock \href
  {http://dx.doi.org/https://doi.org/10.1016/j.neucom.2004.01.019}
  {\path{doi:https://doi.org/10.1016/j.neucom.2004.01.019}}.

\bibitem{Bruno12006}
R.~M. Bruno, B.~Sakmann, Cortex is driven by weak but synchronously active
  thalamocortical synapses, Science 312~(5780) (2006) 1622--1627.
\newblock \href {http://dx.doi.org/10.1126/science.1124593}
  {\path{doi:10.1126/science.1124593}}.

\bibitem{Kumar2010}
A.~Kumar, S.~Rotter, A.~Aertsen, Spiking activity propagation in neuronal
  networks: reconciling different perspectives on neural coding, Nature Reviews
  Neuroscience 11 (2010) 615--627.
\newblock \href {http://dx.doi.org/10.1038/nrn2886}
  {\path{doi:10.1038/nrn2886}}.

\bibitem{Adrian1928}
E.~D. Adrian, The basis of sensation, W W Norton and Co, 1928.
\newblock \href {http://dx.doi.org/10.1016/0166-2236(92)90361-B}
  {\path{doi:10.1016/0166-2236(92)90361-B}}.

\bibitem{Kostal2007}
L.~Kostal, P.~Lansky, J.-P. Rospars, Review article: Neuronal coding and
  spiking randomness, European Journal of Neuroscience 26~(10) (2007)
  2693--2701.
\newblock \href {http://dx.doi.org/10.1111/j.1460-9568.2007.05880.x}
  {\path{doi:10.1111/j.1460-9568.2007.05880.x}}.

\bibitem{Shadlen1998}
M.~N. Shadlen, W.~T. Newsome, The variable discharge of cortical neurons:
  Implications for connectivity, computation, and information coding, Journal
  of Neuroscience 18~(10) (1998) 3870--3896.
\newblock \href {http://dx.doi.org/10.1523/JNEUROSCI.18-10-03870.1998}
  {\path{doi:10.1523/JNEUROSCI.18-10-03870.1998}}.

\bibitem{Gerstner2000}
W.~Gerstner, Population dynamics of spiking neurons: Fast transients,
  asynchronous states, and locking, Neural Comput. 12~(1) (2000) 43--89.
\newblock \href {http://dx.doi.org/10.1162/089976600300015899}
  {\path{doi:10.1162/089976600300015899}}.

\bibitem{Kandel1991}
E.~Kandel, J.~Schwartz, T.~Jessell, Principles of Neural Science, Prentice-Hall
  International edit, Elsevier, 1991.

\bibitem{Brunel1999}
N.~Brunel, V.~Hakim, Fast global oscillations in networks of integrate-and-fire
  neurons with low firing rates, Neural Computation 11~(7) (1999) 1621--1671.
\newblock \href {http://dx.doi.org/10.1162/089976699300016179}
  {\path{doi:10.1162/089976699300016179}}.

\bibitem{Knight2000}
B.~W. Knight, Dynamics of encoding in neuron populations: Some general
  mathematical features, Neural Computation 12~(3) (2000) 473--518.
\newblock \href {http://dx.doi.org/10.1162/089976600300015673}
  {\path{doi:10.1162/089976600300015673}}.

\bibitem{Dayan2005}
P.~Dayan, L.~F. Abbott, Theoretical Neuroscience: Computational and
  Mathematical Modeling of Neural Systems, The MIT Press, 2005.

\bibitem{Masuda2005}
N.~Masuda, B.~Doiron, A.~Longtin, K.~Aihara, Coding of temporally varying
  signals in networks of spiking neurons with global delayed feedback, Neural
  Computation 17 (2005) 2139--2175.
\newblock \href {http://dx.doi.org/10.1162/0899766054615680}
  {\path{doi:10.1162/0899766054615680}}.

\bibitem{Wang2009}
S.~Wang, C.~Zhou, Information encoding in an oscillatory network, Phys. Rev. E
  79 (2009) 061910.
\newblock \href {http://dx.doi.org/10.1103/PhysRevE.79.061910}
  {\path{doi:10.1103/PhysRevE.79.061910}}.

\bibitem{Guo2012}
D.~Guo, C.~Li, Population rate coding in recurrent neuronal networks with
  unreliable synapses, Cognitive Neurodynamics 6~(1) (2012) 75--87.
\newblock \href {http://dx.doi.org/10.1007/s11571-011-9181-x}
  {\path{doi:10.1007/s11571-011-9181-x}}.

\bibitem{van1996}
C.~van Vreeswijk, H.~Sompolinsky, Chaos in neuronal networks with balanced
  excitatory and inhibitory activity, Science 274~(5293) (1996) 1724--1726.
\newblock \href {http://dx.doi.org/10.1126/science.274.5293.1724}
  {\path{doi:10.1126/science.274.5293.1724}}.

\bibitem{Emilio2001}
E.~Salinas, T.~J. Sejnowski, Correlated neuronal activity and the flow of
  neural information, Nature Reviews Neuroscience 2 (2001) 539--550.
\newblock \href {http://dx.doi.org/10.1038/35086012}
  {\path{doi:10.1038/35086012}}.

\bibitem{Brunel2000}
N.~Brunel, Dynamics of sparsely connected networks of excitatory and inhibitory
  spiking neurons, Journal of Computational Neuroscience 8~(3) (2000) 183--208.
\newblock \href {http://dx.doi.org/10.1023/A:1008925309027}
  {\path{doi:10.1023/A:1008925309027}}.

\bibitem{Carsten2003}
C.~Mehring, U.~Hehl, M.~Kubo, M.~Diesmann, A.~Aertsen, Activity dynamics and
  propagation of synchronous spiking in locally connected random networks,
  Biological Cybernetics 88 (2003) 395--408.
\newblock \href {http://dx.doi.org/10.1007/s00422-002-0384-4}
  {\path{doi:10.1007/s00422-002-0384-4}}.

\bibitem{Teramae2007}
J.-n. Teramae, T.~Fukai, Local cortical circuit model inferred from power-law
  distributed neuronal avalanches, Journal of Computational Neuroscience 22~(3)
  (2007) 301--312.
\newblock \href {http://dx.doi.org/10.1007/s10827-006-0014-6}
  {\path{doi:10.1007/s10827-006-0014-6}}.

\bibitem{Kremkow2010}
J.~Kremkow, A.~Aertsen, A.~Kumar, Gating of signal propagation in spiking
  neural networks by balanced and correlated excitation and inhibition, Journal
  of Neuroscience 30~(47) (2010) 15760--15768.
\newblock \href {http://dx.doi.org/10.1523/JNEUROSCI.3874-10.2010}
  {\path{doi:10.1523/JNEUROSCI.3874-10.2010}}.

\bibitem{Kumar2008}
A.~Kumar, S.~Rotter, A.~Aertsen, Conditions for propagating synchronous spiking
  and asynchronous firing rates in a cortical network model, Journal of
  Neuroscience 28~(20) (2008) 5268--5280.
\newblock \href {http://dx.doi.org/10.1523/JNEUROSCI.2542-07.2008}
  {\path{doi:10.1523/JNEUROSCI.2542-07.2008}}.

\bibitem{Vogels2009}
T.~P. Vogels, L.~F. Abbott, Gating multiple signals through detailed balance of
  excitation and inhibition in spiking networks, Nature Neuroscience\href
  {http://dx.doi.org/10.1038/nn.2276} {\path{doi:10.1038/nn.2276}}.

\bibitem{Julien2004}
J.~Mayor, W.~Gerstner, Transient information flow in a network of excitatory
  and inhibitory model neurons: Role of noise and signal autocorrelation,
  Journal of Physiology-Paris 98~(4) (2004) 417 -- 428.
\newblock \href
  {http://dx.doi.org/https://doi.org/10.1016/j.jphysparis.2005.09.009}
  {\path{doi:https://doi.org/10.1016/j.jphysparis.2005.09.009}}.

\bibitem{Han2015}
R.~Han, J.~Wang, H.~Yu, B.~Deng, X.~Wei, Y.~Qin, H.~Wang, Intrinsic
  excitability state of local neuronal population modulates signal propagation
  in feed-forward neural networks, Chaos: An Interdisciplinary Journal of
  Nonlinear Science 25~(4) (2015) 043108.
\newblock \href {http://dx.doi.org/10.1063/1.4917014}
  {\path{doi:10.1063/1.4917014}}.

\bibitem{Xiaojing2017}
J.~Barral, X.-J. Wang, A.~Reyes, Propagation of spike timing and firing rate in
  feedforward networks reconstituted in vitro, bioRxiv\href
  {http://dx.doi.org/10.1101/151134} {\path{doi:10.1101/151134}}.

\bibitem{Neuron-wiki}
{Wikipedia contributors}, \href{https://en.wikipedia.org/wiki/Neuron}{Neuron
  --- {Wikipedia}{,} the free encyclopedia}, [Online; accessed 26-August-2018]
  (2018).
\newline\urlprefix\url{https://en.wikipedia.org/wiki/Neuron}

\bibitem{Shrivastava2011}
A.~Shrivastava, A.~Triller, W.~Sieghart, Gabaa receptors: Post-synaptic
  co-localization and cross-talk with other receptors, Frontiers in Cellular
  Neuroscience 5 (2011) 7.
\newblock \href {http://dx.doi.org/10.3389/fncel.2011.00007}
  {\path{doi:10.3389/fncel.2011.00007}}.

\bibitem{Kantamneni2015}
S.~Kantamneni, Cross-talk and regulation between glutamate and gabab receptors,
  Frontiers in Cellular Neuroscience 9 (2015) 135.
\newblock \href {http://dx.doi.org/10.3389/fncel.2015.00135}
  {\path{doi:10.3389/fncel.2015.00135}}.

\bibitem{Hodgkin1952}
A.~L. Hodgkin, A.~F. Huxley, Currents carried by sodium and potassium ions
  through the membrane of the giant axon of loligo, The Journal of Physiology
  116~(4) (1952) 449--472.
\newblock \href {http://dx.doi.org/10.1113/jphysiol.1952.sp004717}
  {\path{doi:10.1113/jphysiol.1952.sp004717}}.

\bibitem{Hansel1993}
D.~Hansel, G.~Mato, C.~Meunier, Phase dynamics for weakly coupled
  hodgkin-huxley neurons, EPL (Europhysics Letters) 23~(5) (1993) 367.
\newblock \href {http://dx.doi.org/10.1209/0295-5075/23/5/011}
  {\path{doi:10.1209/0295-5075/23/5/011}}.

\bibitem{Wang1996}
X.-J. Wang, G.~Buzs{\'a}ki, Gamma oscillation by synaptic inhibition in a
  hippocampal interneuronal network model, Journal of Neuroscience 16~(20)
  (1996) 6402--6413.
\newblock \href {http://dx.doi.org/10.1523/JNEUROSCI.16-20-06402.1996}
  {\path{doi:10.1523/JNEUROSCI.16-20-06402.1996}}.

\bibitem{Wang2006}
S.~Wang, W.~Wang, F.~Liu, Propagation of firing rate in a feed-forward neuronal
  network, Phys. Rev. Lett. 96 (2006) 018103.
\newblock \href {http://dx.doi.org/10.1103/PhysRevLett.96.018103}
  {\path{doi:10.1103/PhysRevLett.96.018103}}.

\bibitem{Thorpe1996}
S.~I. Thorpe, D.~Fize, C.~Marlot, Speed of processing in the human visual
  system, Nature 381 (1996) 520--522.
\newblock \href {http://dx.doi.org/10.1038/381520a0}
  {\path{doi:10.1038/381520a0}}.

\bibitem{Thorpe2001}
S.~J. Thorpe, A.~Delorme, R.~van Rullen, Spike-based strategies for rapid
  processing, Neural networks : the official journal of the International
  Neural Network Society 14 6-7 (2001) 715--25.
\newblock \href {http://dx.doi.org/10.1016/S0893-6080(01)00083-1}
  {\path{doi:10.1016/S0893-6080(01)00083-1}}.

\bibitem{Rullen2001}
R.~van Rullen, S.~J. Thorpe, Rate coding versus temporal order coding: What the
  retinal ganglion cells tell the visual cortex, Neural Computation 13 (2001)
  1255--1283.
\newblock \href {http://dx.doi.org/10.1162/08997660152002852}
  {\path{doi:10.1162/08997660152002852}}.

\bibitem{Gollisch2008}
T.~Gollisch, M.~Meister, Rapid neural coding in the retina with relative spike
  latencies, Science 319~(5866) (2008) 1108--1111.
\newblock \href {http://dx.doi.org/10.1126/science.1149639}
  {\path{doi:10.1126/science.1149639}}.

\bibitem{Olshausen2004}
B.~A. Olshausen, D.~J. Field, Sparse coding of sensory inputs., Current opinion
  in neurobiology 14 4 (2004) 481--7.
\newblock \href {http://dx.doi.org/10.1016/j.conb.2004.07.007}
  {\path{doi:10.1016/j.conb.2004.07.007}}.

\end{thebibliography}
		
\end{document}